\input harvmac
\baselineskip=16pt
\def \td {\tilde}
\def \const {{\rm const}}

\def\n{\nu}
\def\a{\alpha}
\def\b{\beta}
\def\g{\gamma}

\def\r{\rho}

\def\l{\lambda}
\def\te{\theta}

\def\p{\phi}

\def \y {{\rm y}}
\def \z {{ \rm z}}
\def \f {{\rm f}}
\def \w {{\rm w}}
\def \om {\omega}

\def \ze {\zeta}

\def \ge {{\rm g}}

\def \r {\rho}\def \g {\gamma}  
\def \ov {\over }\def \b {\beta}
\def \P {\Phi}  \def \const {{\rm const}}



\def\mop#1{\mathop{\rm #1}\nolimits}
\def\Li{\mop{Li}}

\lref\jthroat{
J.~Maldacena,
``The large N limit of superconformal field theories and supergravity,''
Adv.\ Theor.\ Math.\ Phys.\ {\bf 2}, 231 (1998)
[hep-th/9711200].
 }

\lref\GKP{
S.~S.~Gubser, I.~R.~Klebanov and A.~M.~Polyakov,
``Gauge theory correlators from non-critical string theory,''
Phys.\ Lett.\ B {\bf 428}, 105 (1998)
[hep-th/9802109].
}

\lref\EW{
E.~Witten,
Adv.\ Theor.\ Math.\ Phys.\ {\bf 2}, 253 (1998)
[hep-th/9802150].
}

\lref \KN{
I.~R.~Klebanov and N.~A.~Nekrasov,
``Gravity duals of fractional branes and logarithmic RG flow,''
Nucl.\ Phys.\ B {\bf 574}, 263 (2000)
[hep-th/9911096].
}

\lref\KT{
I.~R.~Klebanov and A.~A.~Tseytlin,
``Gravity Duals of Supersymmetric $SU(N) \times SU(N+M)$ Gauge Theories,''
Nucl.\ Phys.\ B {\bf 578}, 123 (2000)
[hep-th/0002159].
}

\lref\KW{I.~R.~Klebanov and E.~Witten,
``Superconformal field theory on threebranes at a Calabi-Yau  singularity,''
Nucl.\ Phys.\ B {\bf 536}, 199 (1998)
[hep-th/9807080].
}

\lref\MP{
D.~R.~Morrison and M.~R.~Plesser,
``Non-spherical horizons. I,''
Adv.\ Theor.\ Math.\ Phys.\ {\bf 3}, 1 (1999)
[hep-th/9810201].
}

\lref\Buch{A. Buchel, ``Finite temperature resolution of the 
Klebanov--Tseytlin singularity,'' hep-th/0011146.
}
\lref\KS{I.~R.~Klebanov and M.~J.~Strassler, 
``Supergravity and a Confining Gauge Theory:  
Duality Cascades and $\chi$SB-Resolution of Naked Singularities,''
JHEP {\bf 0008}, 052 (2000)
[hep-th/0007191].
}
\lref\PT{L. A.  Pando Zayas and A. A. Tseytlin,
``3-branes on spaces with $R \times  S^2 \times  S^3$ topology,''
hep-th/0101043.
}
\lref\PTtwo{L.~A.~Pando Zayas and A.~A.~Tseytlin,
``3-branes on resolved conifold,''
JHEP {\bf 0011}, 028 (2000)
[hep-th/0010088].
}

\lref\SK{A.~Kehagias and K.~Sfetsos,
``On running couplings in gauge theories from type-IIB supergravity,''
Phys.\ Lett.\ B {\bf 454}, 270 (1999)
[hep-th/9902125].
``On asymptotic freedom and confinement from type-IIB supergravity,''
Phys.\ Lett.\ B {\bf 456}, 22 (1999)
[hep-th/9903109].
}
\lref\Gub{S.~S.~Gubser,
``Dilaton-driven confinement,''
hep-th/9902155.
} 
\lref\gim{G.~W.~Gibbons and K.~Maeda,
``Black Holes And Membranes In Higher Dimensional Theories With Dilaton Fields,''
Nucl.\ Phys.\ B {\bf 298}, 741 (1988).
}
\lref\dup{  M.~J.~Duff, H.~Lu and C.~N.~Pope,
``The black branes of M-theory,''
Phys.\ Lett.\ B {\bf 382}, 73 (1996)
[hep-th/9604052].
 } 
\lref\call{ C.~G.~Callan, S.~S.~Gubser, I.~R.~Klebanov and A.~A.~Tseytlin,
``Absorption of fixed scalars and the D-brane approach to black holes,''
Nucl.\ Phys.\ B {\bf 489}, 65 (1997)
[hep-th/9610172].
}
\lref \HS {G.~T.~Horowitz and A.~Strominger,
``Black strings and P-branes,''
Nucl.\ Phys.\ B {\bf 360}, 197 (1991).
}
\lref\Cvetic{
M.~Cvetic, H.~Lu and C.~N.~Pope,
``Brane resolution through transgression,''
hep-th/0011023.
M.~Cvetic, G.~W.~Gibbons, H.~Lu and C.~N.~Pope,
``Ricci-flat metrics, harmonic forms and brane resolutions,''
hep-th/0012011.
}
\lref \HK {C.~P.~Herzog and I.~R.~Klebanov,
``Gravity duals of fractional branes in various dimensions,''
hep-th/0101020.
}
\lref\bhkpt{A. Buchel, C. Herzog, I.R. Klebanov, L. Pando Zayas and
A.A. Tseytlin, ``Non-Extremal Gravity Duals of Fractional D3-branes
on the Conifold,'' hep-th/0102105.
}

\lref\gkp{
S.~S.~Gubser, I.~R.~Klebanov and A.~W.~Peet,
``Entropy and Temperature of Black 3-Branes,''
Phys.\ Rev.\ D {\bf 54}, 3915 (1996)
[hep-th/9602135].
}

\lref\Witt{
E.~Witten,
``Anti-de Sitter space, thermal phase transition, and confinement in  gauge theories,''
Adv.\ Theor.\ Math.\ Phys.\ {\bf 2}, 505 (1998)
[hep-th/9803131].
}

\lref\krasnitz{M. Krasnitz, ``A two-point function in a cascading
${\cal N}=1$ gauge theory from supergravity,'' hep-th/0011179.
}

\Title{\vbox
{\baselineskip 10pt
{\hbox{PUPT-1978}
\hbox{OHSTPY-HEP-T-01-003}
\hbox{NSF-ITP-01-15}
\hbox{CALT-68-2317}
\hbox{CITUSC-00-069}
\hbox{hep-th/0102172}
}}}
{\vbox{\vskip -30 true pt\centerline { 
Restoration of Chiral Symmetry: }
\medskip
\centerline { A Supergravity Perspective  }
\medskip
\vskip4pt }}
\vskip -20 true pt
\centerline{S. S. Gubser$^{1,2,3}$,
C. P. Herzog$^{3}$, I. R.~Klebanov$^{1,3}$ 
and A. A. Tseytlin$^{4,5}$\footnote{$^*$} {Also at Lebedev 
Physics Institute, Moscow. }
 }
\smallskip\smallskip
\centerline{$^{1}$ \it Institute for Theoretical Physics,
University of California, Santa Barbara, California 93106-4030}
\centerline{$^{2}$ \it California Institute of Technology 452-48,
Pasadena, California 91125}
\centerline{$^{3}$ \it Joseph Henry
Laboratories, Princeton University, Princeton, New Jersey 08544}
\centerline{$^{4}$ \it  Department of Physics, 
The Ohio State University,  
Columbus, OH 43210 }
\centerline{$^{5}$ \it  Blackett Laboratory, 
Imperial College,   London SW7 2BZ, U.K. }

\bigskip\bigskip
\centerline {\bf Abstract}
\baselineskip12pt
\noindent
\medskip
The supergravity dual of
$N$ regular and $M$ fractional D3-branes on the conifold
has a naked singularity in the infrared. Supersymmetric resolution
of this singularity requires deforming the conifold: this is the supergravity
dual of chiral symmetry breaking. Buchel suggested that at sufficiently
high temperature there is no need to deform the conifold: the singularity
may be cloaked by a horizon. 
This would be the supergravity manifestation of chiral symmetry restoration.
In previous work [hep-th/0102105] the ansatz and the system
of second-order radial differential 
equations necessary to find such a solution were
written down. In this paper we find smooth solutions to this system
in a perturbation theory that is valid when the Hawking temperature
of the horizon is very high.
\bigskip

\Date{02/01}

\noblackbox 
\baselineskip 15pt plus 2pt minus 2pt

\newsec{Introduction}

The AdS/CFT correspondence \refs{\jthroat,\GKP,\EW}
has produced a wealth of new information
about strongly coupled conformal gauge theories.
Considerable effort has also been invested into extending it
to non-conformal theories. One recent development is
the supergravity  description  of chiral symmetry breaking
in a certain ${\cal N}=1$ supersymmetric $SU(N)\times SU(N+M)$
gauge theory \KS. This theory may be realized by
adding $M$ fractional D3-branes (wrapped D5-branes) to
$N$ regular D3-branes at the apex of the conifold, which is
defined by the constraint $\sum_{i=1}^4 z_i^2=0$ in
${\bf C}^4$.
For $M=0$ this gauge theory reduces to the superconformal theory
dual to the $AdS_5\times T^{1,1}$ background of type IIB string 
\refs{\KW,\MP}.

In the supergravity dual the $M$ fractional branes
are replaced by $M$ units of RR 3-form flux through the
3-cycle of the compact space. This flux changes the background and
introduces the logarithmic running of $\int_{S^2} B_2$, which is related
to the running of field theoretic couplings \KN.
In turn, this causes the RR 5-form flux to 
grow logarithmically with the radius \KT, 
due to the equation $dF_5 = H_3\wedge F_3$.
In \KS\ this behavior was attributed to a cascade of Seiberg
dualities in the dual ${\cal N}=1$ supersymmetric $SU(N)\times SU(N+M)$
gauge theory.

While the solution of \KT\ is smooth in the UV (for large $\r$),
it has a naked singularity in the IR. To resolve this singularity
while preserving the ${\cal N}=1$ supersymmetry, it is necessary
to deform the conifold \KS, i.e.\ to replace the constraint with
$\sum_{i=1}^4 z_i^2=\epsilon^2$. 
The resulting solution, a warped deformed conifold,
is perfectly non-singular and
without a horizon in the IR, while it asymptotically approaches
the KT solution \KT\ in the UV.
The mechanism that removes the naked singularity is related to the
breaking of the chiral symmetry in the dual $SU(N)\times SU(N+M)$
gauge theory. The ${\bf Z}_{2M}$ chiral symmetry, which may be approximated by
$U(1)$ for large $M$, is realized geometrically as
$z_i\to z_i e^{i\theta}$. The deformation of
the conifold breaks it down to ${\bf Z}_2$: $z_i\to -z_i$ \KS.

In \Buch\ a different mechanism for resolving the KT naked singularity
was proposed. It was suggested that a non-extremal generalization of
the KT solution, with unbroken $U(1)$ symmetry, may have a regular
Schwarzschild horizon ``cloaking'' the naked singularity. The dual
field theory interpretation of this would be the restoration of chiral
symmetry above some critical temperature $T_c$ \Buch. 
 The proposal of \Buch\ is that the description of the phase with
restored symmetry involves a regular Schwarzschild horizon appearing
in the asymptotically KT geometry. This proposal is analogous to the
fact that the ${\cal N}=4$ SYM theory, which is not confining, is
described at a finite temperature by a black hole in $AdS_5$ 
\refs{\gkp,\Witt}.

We may ask, to what extent can chiral symmetry breaking and
confinement be studied, for fractional D3-branes on the conifold, via
black hole horizons?  It is the aim of this paper to demonstrate that
at least the high temperature phase is accessible (thus realizing the
goal of \Buch).  The thermodynamics of
the low temperature phase might be difficult to treat
from a classical supergravity perspective.  The reason is that at low
temperatures, where confinement has occurred, there are $O(N^0)$
degrees of freedom, whereas on the classical supergravity side, a
horizon has an entropy of $O(N^2)$, where $N$ is the relevant number
of colors.  So it would seem that to study the thermodynamics of the
$T < T_c$  phase we need to go at least to one-loop effects on
the string theory side.

We leave open the question of how well one might describe in
supergravity terms the phase transition where chiral symmetry is
broken.  Standard lore suggests that this is a second order transition
at finite $T_c$.  The geometry dual to the theory at criticality
should admit some action of the conformal group; or at the least, some
correlation functions, computed with periodic Euclidean time, should
have power law tails in the infrared.  Now, the latter statement is
impossible as long as there is a smooth horizon, because in the
periodic Euclidean time formalism, 
a smooth horizon means that the non-compact space has the topology
of a disk (in the $t$-$r$ directions)
times ${\bf R}^3$, and the disk factor will invariably generate a gap,
as observed in \Witt.
  The alternative is to have the topology of a
cylinder times ${\bf R}^3$.  This would be the case, for instance, if
the geometry dual to the critical point were Euclidean $AdS_5$ with
time made periodic.  Such a geometry is the obvious candidate to
describe criticality,\foot{It seems difficult however to find an
anti-de Sitter solution of supergravity with the fluxes appropriate to
fractional D3-branes.} since it does admit an obvious action of the
spatial part of the conformal group, which can be preserved at finite
temperature.  The horizon is degenerate, and by the usual rules of
black hole thermodynamics has no entropy---which is only to say that
the entropy is subleading in $N$.  The same statements seem to apply
whenever the topology in the $t$-$r$ direction is a cylinder.

The considerations of the previous paragraphs all tend toward the
conclusion that a regular black hole horizon should appear only at
some finite Hawking temperature, and that this temperature should be
the $T_c$ of chiral symmetry breaking.  For temperatures below $T_c$,
we might expect non-extremal generalizations of the KS solution which
are free of horizons, just like the extremal solution.  
The absence of a horizon in the extremal geometry results in an
area law for Wilson loops, which is a manifestation of confinement.
Near-extremal generalizations of it should retain this area-law
property.
Clearly it would be very interesting to study the $T=T_c$
point in supergravity, to the extent that this is possible.  We
suggest a limiting procedure which may be used for finding the
appropriate supergravity background.

In order to address physically interesting questions at finite
temperature from a dual supergravity point of view, we need to study
non-extremal generalizations of the KT and KS backgrounds.  This
problem was first addressed for the KT background by Buchel in \Buch\
partly with numerical methods.  However, in \bhkpt\ this solution was
shown to be singular.  It was argued  there 
that a more general ansatz where
the 3-forms are not  self-dual is necessary to find a
regular Schwarzschild horizon.

In this paper we show that solutions of this type indeed exist.
First we review the ansatz and the basic equations derived in
\bhkpt. Then we proceed to develop perturbation theory that
is valid in the high temperature phase $T \gg T_c$.

\newsec{Non-Extremal Generalization of the KT Ansatz}

We start with the ansatz of \bhkpt\ for the non-extremal KT
background describing the high temperature phase
where the chiral symmetry is restored.
The 10-d  Einstein-frame metric was taken to be of the general
form
 consistent with the $U(1)$ symmetry of $\psi$-rotations
and the interchange of the two $S^2$'s. It involves 
4 functions $x,y,z,w$  of a radial coordinate $u$:
\eqn\mett{
ds^2_{10E} =  e^{2z} ( e^{-6x} dX_0^2 + e^{2x} dX_i dX_i)
+ e^{-2z}  ds^2_6 \ ,   }
where 
\eqn\mott{ds^2_6 =  e^{10y} du^2 + e^{2y} (dM_5)^2  \ , } 
\eqn\mmm{
(dM_5)^2 = e^{ -8w}  e_{\psi}^2 +  e^{ 2w}
\big(e_{\theta_1}^2+e_{\phi_1}^2 +
e_{\theta_2}^2+e_{\phi_2}^2\big) \equiv
 e^{ 2w} ds_5^2    \ , }
and 
$$
 e_{\psi} =  {1\ov 3} (d\psi +  \cos \theta_1 d\phi_1  +  \cos \theta_2 d\phi_2)  \  , 
 \quad  e_{\theta_i}={1\ov \sqrt 6} d\theta_i\ ,  \quad  e_{\phi_i}=
{1\ov \sqrt 6} \sin\theta_id\phi_i \ .
$$
Here $X_0$ is the euclidean time and $X_i$ are the 3 longitudinal 3-brane directions.

This metric  can  be brought into a more familiar D3-brane form
\eqn\fop{
ds^2_{10E} =  h^{-1/2}(\r)  [ g(\r)  dX_0^2 + dX_i dX_i]
+  h^{1/2}(\r)   [  \ge^{-1} (\r) d\r^2 
+ \r^2  ds^2_5] \ , } 
with the redefinitions 
\eqn\iop{
 h=  e^{-4z- 4x} \ , \ \ \ \ \     \r  = e^{y + x + w  } ,
 \ \ \
\ \ \     g=  e^{-8x}\  ,\ \ \ \ \ \ \ e^{10y + 2x} du^2  = \ge^{-1} (\r) d\r^2 \ .  }
When $w=0$ and $e^{4y}=\rho^4= { 1 \ov 4u} $,
the transverse 6-d space is the standard conifold
with $M_5= T^{1,1}$.  
Small  $u$  thus  corresponds to large distances
(where we shall assume that $g,{\rm g}$
and also $h $, in the asymptotically flat  case, 
approach 1 
  as $\r \to \infty$)
 and vice versa.
The function $w$ squashes
the $U(1)$ fiber of $T^{1,1}$ relative to the 2-spheres; it does not
violate the $U(1)$ symmetry.

The ansatz for the $p$-form  fields is dictated by 
symmetries  and thus  is  exactly the same 
as in the extremal KT case \KT:\foot{Note that
 the function $T$ in \KT\ is  related to $f$ used in \bhkpt\ and here by 
 $f= { 1 \ov \sqrt 2 } T$, and we make a similar rescaling of $P$:
$P = { 1 \ov \sqrt 2 }  P_{KT} $.
}
\eqn\har{
F_3 = \    P
e_\psi \wedge
( e_{\theta_1} \wedge e_{\phi_1} - 
e_{\theta_2} \wedge e_{\phi_2})\ ,
} 
\eqn\ansa{
B_2  = \    f(u) 
( e_{\theta_1} \wedge e_{\phi_1} - 
e_{\theta_2} \wedge e_{\phi_2})
 \ , 
}
\eqn\fiff{
F_5= {\cal F}+*
{\cal F}\
, \quad  \ \ \ \ {\cal F} =  K(u) 
e_{\psi}\wedge e_{\te_1} \wedge
e_{\p_1} \wedge
e_{\te_2}\wedge e_{\p_2}\ . }
As in \KT, the  Bianchi identity for the
5-form, \ $
d*F_5=dF_5=H_{3}\wedge F_3$,  implies
\eqn\kee{
K (u)  = Q + 2 P f (u) \ .}
We will not generally impose the self-duality of the 3-forms,
which means that we have to include the dynamics of the dilaton $\Phi$
and of the ``squash factor'' $w$.

\newsec{Basic System of Equations and its
Limiting Solutions }

The system of radial equations that need to be solved was derived
in \bhkpt. 
The simplest equation is  for the ``non-extremality function'' $x$
\eqn\bass{ x''=0 \ ,\ \ \ {\rm i.e.} \ \  \ 
 x= a u \ , \ \ \ \ a =\const \ , }
and in the non-extremal case we take $a > 0 $.
The coupled equations for the remaining five   
functions   of $u$, i.e. 
$y, z, w,  \P$  and $f$ or, equivalently, $K$, 
   are 
\eqn\yyy{ 10y'' - 8 e^{8y} (6 e^{-2w} - e^{-12 w})   + \P''
=0 \ , 
}
\eqn\yuw{
10w'' - 12 e^{8y} ( e^{-2w} - e^{-12 w})   - \P''
=0 \ , } 
\eqn\ppp{
\P''    + e^{-\P + 4z - 4y-4w} 
\left ({K'^2\over 4P^2} -  e^{2 \P + 8 y+8w} P^2 \right )=0 \ , }
\eqn\zzy{
4z'' -  K^2  e^{8z}
 - e^{-\P + 4z - 4y-4w} \left ({K'^2\over 4P^2}
    +  e^{2 \P + 8 y+8w} P^2 \right ) =0\ , 
}
\eqn\fef{
(e^{-\P + 4z - 4y-4w} K')' - 2P^2 K e^{8z} =0 \ . 
}
 The  integration constants are 
subject  to the  zero-energy constraint  $T + V =0$, i.e.
\foot{The  above system  follows from the 1-d effective lagrangian  $L= T-V$.} 
$$5  y'^2   - 2 z'^2  - 5 w'^2 - { 1 \ov 8} \P'^2 
-   e^{-\P +  4z -4y - 4 w } {K'^2\ov 16 P^2}$$ 
\eqn\coln{ 
 -   \  e^{8y} ( 6 e^{-2w} - e^{-12 w} )
 +  { 1 \ov 4} e^{\P+  4z + 4y + 4 w } P^2 + 
 { 1 \ov 8}  e^{8z} K^2   = 3 a^2  \ . 
}
A note about the dimensions. From the form of  
the metric \mett\ it is natural to require  that 
  $e^y$   and  $u^{-1/4}$   should 
 have dimension of length, while $x,z,w$ 
should be  dimensionless.  Since we have set the 10-d 
gravitational constant   to be 1 
(i.e. we measure the scales in terms of the 
10-d ``Planck scale'' $L_{\rm P}\sim (g_s \a'^2)^{1/4}$),\foot{
Since we are concerned with gauge/gravity duality in this paper,
$L_{\rm P}$ has nothing to do with the physical Planck length. Instead,
due to the dimensional transmutation in this theory \KS, $L_{\rm P}/P$
sets the scale of glueball masses.}
 then from the 
1-d action  or \coln\  we  conclude  
that  $K$  and $Q$ in \kee\  have dimension  (length)$^4$ 
while  $P$  and $f$  have dimension    (length)$^2$. 
This should be kept in mind  in what follows. 
It is easy to restore the dependence on the  Planck  length 
by rescaling $ Q \to L_{\rm P}^4 Q$, \ $ P \to L^2_{\rm P}  P$, etc. To restore 
the dependence  on the string coupling 
one should further   rescale $P^2 \to g_s P^2$. At the 
end,  $ Q \sim g_s  \a'^2  N  , \  P \sim g_s \a' M $.

As shown in \refs{\Buch,\bhkpt},  there is a  subclass of  simple  non-extremal 
solutions for which the functions $K$ and $z$
satisfy the same 1-st order equations as in the extremal case \KT.  Indeed, 
if we set 
$ K'= -  2 P^2 e^{ \P + 4 y+4w} $, then \fef\ 
implies that $z$ should be subject to a 1-st order equation. 
In this case the 3-forms are self-dual.
Then it is consistent, in particular, 
 to keep $\Phi=0$ and $w=0$ so that $T^{1,1}$ is not squashed.
However, these simplest solutions turn out to be
 singular: they have a horizon coinciding  with a curvature 
singularity \bhkpt.

To  find a 
non-extremal generalization of the KT solution \KT\ with a 
{\it regular}  horizon  of the Schwarzschild type
it is necessary to study the  above second  order system in
its  full generality.
We shall be looking  for a solution 
 which  satisfies the two natural requirements: 

\noindent
  (a) it is  a one-parameter  ($x'=a$ or  Hawking temperature) 
generalization of  the  extremal   KT solution;

\noindent
(b) it reduces to the standard {\it regular}  black D3-brane  solution 
  in the $P=0$ limit.
 
\noindent
Thus  for $a\to 0$ the solution should reduce to the KT one \KT\ 
\eqn\kot{x=w=\P=0\ , \ \ \ \ \ \ 
  e^{-4y} = 4u \ , \ \  \ \ \ \ \ \ 
\ K= - {P^2\ov 2} \log (u L_{\rm P}^4) \ ,   } 
 \eqn\ktt{ h= e^{-4z} = h_0    
 -  {P^2\ov 2}  u [ \log (u L_{\rm P}^4) -1 ] \ ,  \ \ \ \ \ \ \   
 }
where $h_0=1$ for the asymptotically flat solution 
and $h_0=0$ for the analog of the near-horizon AdS part of  
the pure D3-brane background.

At the same time, for 
  $P\to 0$  the solution should  reduce to  
  the regular black  D3-brane background 
\eqn\aau{ w=\P=0\ , \ \  \ \ \  e^{4x} = e^{4au}  \ ,  
\ \ \ \ \ \ e^{-4y} =  { a^{-1} \sinh 4a u  } \ , \ \ \ \ \ 
 e^{-4z} =  {   {Q\over 4 a} \sinh 4 a (u + k)  } 
\ .  
}
For the asymptotically flat ($h(0)=1$)
 boundary conditions\foot{The
choice of $k=0$ in \aau\
leads to the  black hole in AdS  solution, 
where  we drop the asymptotically flat region, 
i.e. $h(0)=0$.  
In this case 
 $\g=1$, i.e. $\td Q=Q$ in the expression for $h$ below.} 
\eqn\iu{
  e^{ 4a k}=\g = Q^{-1}
 ( \sqrt{ Q^2 + 16 a^2} + 4a ) \ ,    
}
the  metric  \fop\  should take the 
standard non-extremal D3-brane form \refs{\HS,\dup} 
with 
\eqn\hji{  g=\ge= e^{-8x} = 1 - {2 a\ov  \r^{4}}  \  , 
 \ \ \ \ \ \ \ \ \ \ 
  \r^4= e^{ 4y+4x} = { 2 a \ov 1- e^{-8a u} } \ , }
\eqn\qqq{
h=  e^{-4z- 4x}= 1 + { \tilde Q \ov 
 4\r^{4} }  \ ,  \ \
 \ \ \ \ \ \ \ \
 \tilde Q = \g^{-1} Q  
=  \sqrt{ Q^2 +  16 a^2 }  - 4 a   \  .  }
Note that  near the horizon ($u\to \infty$):
\eqn\eess{ 
y=  - a u + { \log 2 a \ov 4}  +    {  e^{-8au} \ov 4}
  + O( e^{-16au}) \ , \ \ \ \ \
z=  - a u +   { \log {8 a\ov Q \g} \ov 4} 
+    {  e^{-8au} \ov 4\g^2}   + O( e^{-16au}) \ ,
}
 while at large distances   ($u\to 0$)
\eqn\sss{
\  y= - { 1 \ov 4} \log 4 u  - { 2 \ov 3} a^2 u^2 + O(u^3)  
\ , \ \ \ \ \  \ \ \ \ 
e^{-4z}= {Q\ov 8a} ( \g - \g^{-1} ) 
+ {Q\ov 2} ( \g +  \g^{-1} ) u +  O( u^2) \ , }
i.e. 
\eqn\sis{
e^{-4z}
= 1 + \sqrt{Q^2 + a^2}\   u  + O( u^2) \ , \  \ \ \ \ \ \ \ \ 
(e^{-4z})_{k=0} =  Q u + O( u^2) \ .}

\newsec{Asymptotics of the Regular Non-Extremal Solution}

Since we do not see a simple analytic solution
to \yyy--\fef\  which has the required properties, 
we need to outline
a strategy for finding it in perturbation  theory or numerically.
This involves understanding the behavior of the solutions in
the two asymptotic regions: $u\rightarrow \infty $ and $u\rightarrow 0$, 
i.e. in the short-distance and long-distance limits.

In order for the  non-extremal generalization of the 
KT solution  to reduce to the standard 
black D3-brane in the $P\to 0$ limit,
the $u\to \infty $  asymptotics   have to  match \eess:  
\eqn\boun{ 
x = a u \ , \ \ \ \   \ \ \ y\to - a u  + y_* \  , \ \  
\ \  \  
z\to - au +  z_*\ 
 ,    }
\eqn\bam{   w\to w_* \ , \ \  \ \ \ \   
\P\to \P_* \ ,  \ \  \ \ \ \ K\to  K_* \ . }
The asymptotics \boun\ guarantee the existence of a regular
Schwarzschild horizon at $u=\infty$, and it
 is natural to expect 
that $w, K$ and $\P$  have stationary points at this 
 horizon.
Then it is easy to see that  our system of equations 
\yyy-\fef\ {\it and} the constraint 
\coln\  are indeed satisfied at large $u$. 
We will also show that turning
on $P$ makes a small perturbation on the large $u$ asymptotics.

 The information on the Hawking temperature
is contained in the constants $y_*$ and $z_*$.
The natural near-horizon variable is $U = e^{-4au}$. Insisting that there is
no conical singularity in the $U-X_0$ plane fixes the  Hawking temperature
to be
\eqn\haw{ T =  { 2 \ov \pi}  a e^{2 z_*- 5 y_*}
\ .  } 
In general, there are 
 different possibilities for the large $u$ 
behaviour: 

\noindent
(i) The asymptotic $u\to \infty $  solution corresponds to the
existence of a regular Schwarzschild horizon, i.e. a solution
with the form \boun.\  

\noindent
(ii) 
$e^{-4z}$   and thus $h= e^{-4z-4x}$ in \fop\ 
 vanish at some {\it finite} $u$ 
before we reach  $u=\infty$; in this case there is a naked
singularity which is not cloaked by a horizon.

\noindent
(iii) It may  also happen that 
 $h= e^{-4z-4x}$
vanishes at $u=\infty$, so we get a
 horizon coinciding with the singularity as in  
the solution of \Buch.

We believe that  possibility (ii) corresponds to $T<T_c$ where
a $U(1)$ symmetric solution is singular and one needs an appropriate
KS-type  ansatz \KS\ to remove the singularity. 
Possibility (i) corresponds to $T>T_c$ where
our ansatz should be sufficient.

At large distances 
($u\to 0 $) the non-extremal solution should approach the extremal
KT solution \kot,\ktt, i.e. 
we require that  
\eqn\lla{
u\to 0: \ \ \ \ \
x,w,\P\ \to\  0 \ , \ \ \  \ \ \ \ 
y \to  - { 1 \ov 4} \log 4u\  .
}
Note that this behavior is also in agreement
with the asymptotics \sss\ found for the regular D3-brane.
The behaviors of the effective 3-brane charge \kee\ 
and of the warp factor are required to be
 \eqn\Kasympt{ K(u) \to -{P^2\over 2} \log (u L_{\rm P}^4)\ ,
 \qquad e^{-4z} \to -  
{P^2\ov 2} u\log (u L_{\rm P}^4)
 \ ,
}
where we adopt the choice $h_0=0$ such that there is no
asymptotically flat region.

If we 
attempt  integrating our 2nd-order equations from $u=0$
starting with these asymptotics, 
we should vary the temperature \haw\ 
 to find solutions
where  possibility (i) is realized and we reach
the large $u$ behavior \boun. 

As we integrate towards large $u$, the effective 3-brane charge
$K(u)$ decreases: this is the varying flux phenomenon.
For low temperature $K(u)$ reaches zero before the non-extremality
effects have a chance to affect the solution significantly. This produces  possibility (ii).
However, we believe that there is a critical temperature $T_c$
above which $K(u)$ is positive everywhere and  possibility (i) is realized.

Thus, the task of the numerical work is to determine the
accessible values of $z_*$ and $y_*$ as a function of the parameters,
and to show that the allowed values of the temperature 
\haw\ are bounded
from below in this symmetry restored phase.

If  possibility (i) is indeed realized, then
we expect  the decreasing $K(u)$ to stabilize
at a value $K_*$ for large $u$.
 If the horizon has the Hawking temperature $T$,
then in the dual field theory we interpret $K_*$ as the effective number
of unconfined color degrees of freedom at temperature $T$.
As $T$ approaches $T_c$ from above, 
$K_*$ should decrease.
This would agree with the dual field theory interpretation: 
as we lower the temperature we excite 
a smaller effective number of colors.
We believe that 
$K_*$ should approach {\it zero} at the critical temperature $T_c$. From 
the field theoretic point of view, $K_* \to 0$ because there is a
reduction in the effective number of degrees of freedom at the phase 
transition.

On the supergravity side,
we can see the special role of the point $K_*=0$ from 
eq. \fef. Using \boun\ we find 
that, if both $dK\ov dU$ and $K$ vanish at the horizon $U=e^{-4au}=0$,
then  ${d^2 K\ov dU^2}=0$ so that $K=0$ everywhere.
Thus, in 
the symmetry-restored phase $K_*$ must be positive. 
As we approach $T_c$ from above, $K_*\rightarrow 0$.
In this limit $K(u)$ becomes very
close to zero in the IR (near the horizon). 

\newsec{ Perturbation Theory in $P$ }

One  useful  approach  to constructing 
the required regular non-extremal solution 
is to   start with the 
 non-extremal D3-brane solution \aau,
 which is valid for
$P=0$, and find its deformation order by  
order in $P^2$.
A remarkable feature of perturbation theory in $P^2$ 
near the extremal ($a=0$)  D3-brane  background is that 
  already the {\it first-order}  correction 
gives  the {\it exact} form of the 
 KT solution \kot,\ktt.
Therefore, it is natural to expect that  a similar 
expansion  near the non-extremal  D3-brane  solution
will capture the basic features of  
non-extremal generalization of the KT background.

More precisely, our starting point will be 
the well-known D3-brane solution \aau\  with $Q$ replaced by the 
effective 3-brane charge $K_*$, so that we
automatically  match onto 
the near-horizon asymptotics \boun,\bam.
Perturbing in $P^2$ around the near-extremal
3-brane solution of charge $K_*$,
we will see that the $O(P^2)$ modification is already enough 
to match onto the 
KT long-distance asymptotics. 
 The  small parameter governing this expansion 
is actually  the dimensionless ratio 
$P^2
K_*^{-1} $, \foot{Note that 
$P\sim g_s M$ and $K\sim g_s N$ where $M$ and
$N$ are the numbers of fractional and regular D3-branes respectively.}
i.e. for this method to work the horizon value of the effective
3-brane charge $K_*$ has to be sufficiently large.  In view of the
discussion in section 4, this means that this perturbation theory is
applicable for $T\gg T_c$.

It is useful to rescale  the
fields  by 
appropriate powers of $P^2$, setting  
\eqn\seeq{
K(u)= K_* + 2 P^2 F(u)
 \ , \ \ \ \ \ \P (u) = P^2 \p(u) \ , \ \  \ \ w(u) = P^2 \om (u) \ ,}
and 
\eqn\ded{y\to   y  +  P^2 \xi\ , \ \ \ \ \  
e^{-4z} \to  e^{-4z}  + P^2  \ze \ ,  \ \ \  
{\rm i.e.} \ \ \
 z\to   z  +  P^2 \eta\ ,  \ \ \ \  \ze
= -4 e^{ - 4 z} \eta  + O(P^2) \ ,    }
where  $y, z$  represent 
 the pure D3-brane solution \ktt :
$ e^{-4y} = a^{-1} \sinh 4au,\ \ e^{-4z} =
 {K_*\ov 4a}\sinh 4a(u+k)$,  and $\xi$ and $\zeta$
or $\eta$
are  corrections to it.
To match onto the small $u$ KT asymptotics,
 $e^{-4z} \to - {P^2\ov 2} u \log (u L_{\rm P}^4) $,   
we require that
\eqn\matching{
\omega (0) = \xi(0) = \p (0) = 0\ , \  \quad \  \ 
F \to -{1\over 4} \log u \ , \ \ \ \ \ 
\zeta \to -{1 \over 2} u
\log (8 a u)\ .}
For $k=0$, i.e. for the case without an asymptotically flat region,  this means
\eqn\match{ 
\eta \to  {1\over 8 K_*} \log (8 a u) \ . }
Note that $\eta$ is not  uniformly small and it seems better to consider
the expansion  in terms of  $\zeta$ which goes to zero.
Moreover, the leading-order correction to $\zeta$
already reproduces the exact KT solution. 
However, 
  $\zeta$  and $\eta$  are  directly related,
so the two expansions are, in fact, equivalent
 and we will find it more convenient  to use $\eta$.

Now the system \yyy--\fef\
takes the following explicit form:
\eqn\yyu{ 10\xi'' - 320  e^{8y} \xi 
   +  \p''  +O(P^2)    =0 \ , }
\eqn\typs{ 
10\om''  -  120 e^{8y} \om - \p'' +   O(P^2)  =0
\ , }
\eqn\pu{
\p'' + e^{ 4z - 4y} (F'^2 - e^{  8 y}) + O(P^2)=0 \ , }
\eqn\fefh{
(e^{ 4z - 4y} F')' -  K_* e^{8z}  + 
 O(P^2)=0 \ . 
}
\eqn\zy{
4\eta'' - 8 K_*^2  e^{8z}\eta 
 -  4  K_*  F e^{8z}
 -    e^{ 4z - 4y} ( F'^2 +  e^{ 8 y})
 +O(P^2) =0\ . 
}
The constraint \coln\ becomes 
\eqn\cvo{
10 y' \xi'  - 4 z' \eta' 
- { 1 \ov 4}   e^{ 4z - 4y}  F'^2
- 40 e^{8 y} \xi 
+ { 1 \ov 4}   e^{ 4z + 4y}
+    K_*^2 e^{8 z}  \eta   + { 1 \ov 2} K_* e^{8 z} F
+ O(P^2) = 0 \ . }

\subsec{\bf Leading-order solution for $K$ }

Using \zzy, i.e.  $ K_* e^{8z} = 4 K_*^{-1} z''$,  
we get  from \fefh,\aau\
\eqn\fre{
F' = c -  K_*^{-1} e^{ 4y} ( e^{-4z})' 
= c  -  { a \cosh 4a(u+k)  \ov   \sinh 4 a u } \ . } 
For large $u$ (near the horizon), 
we must have $ F' \to 0$ in order to satisfy 
\bam.
This fixes  the integration  constant to be 
$c =a \g = a e^{ 4ak} = a \g $,  so that  
\eqn\ffe{
F' = - { 2a \b  \ov e^{8au} -1 } \ , 
\ \ \  \ \ \ \ \ \  \ 
\b =\cosh 4 a k = 
 ( 1 + { 16 a^2\ov K_*^2} )^{1/2}  \ , } 
and thus 
\eqn\exx{ F = -    { 1 \ov 4} \b
 \log 
( 1- e^{-8 a u} ) \ . }
As required by \seeq, this expression satisfies
$F(u\to \infty)\equiv F_* = 0$. 
 
We will be particularly interested in the case $k=0$ 
in \aau,   when 
the starting point of the perturbation theory has AdS asymptotics
rather than joining onto asymptotically flat space.
In this limit $\b = 1$ in \ffe\ \foot{In the 
$k=0$ case the r.h.s. of the expressions in \iu\ and \ffe\
do not, of course,  apply.}
  and thus 
\eqn\samolet{
K(u)=\  K_* -{P^2\over 2} \log ( 1- e^{-8 a u} )
\ .}
This expression approaches $K_*$ for large $u$, as required.
If $P^2 K_*^{-1}  \ll 1$ then the second term is a small perturbation
for almost all values of $u$, except close to zero. Thus, it seems
that our perturbation theory breaks down near $u=0$.
Nevertheless, we note that \samolet\ has precisely the same small
$u$ asymptotics \Kasympt\ as the extremal KT solution!

 Thus, 
already at the leading order 
this perturbation theory produces a solution with
the correct KT asymptotics. This remarkable fact strengthens our 
confidence that an exact solution interpolating between the
KT solution at small $u$ and the regular D3-brane horizon at large $u$
indeed exists. Our perturbed solution should be a good approximation to
it provided that $P^2 K_*^{-1} \ll 1$.  
This limit corresponds to
high Hawking temperatures. To show this,
let us match \samolet\ with \Kasympt\ for small $u$.
We find that
\eqn\hihi{ 8 a L_{\rm P}^{-4} = e^{ 2/\l} \ , \ \ \ \ \ \  \ \ \ 
\l\equiv  {P^2 K_*^{-1} } \ll 1   
\ .}
On the other hand, the Hawking temperature is determined in terms of the
non-extremality $a$ and the charge near horizon $K_*$ by
the usual near-extremal D3-brane formula (cf. \haw,\aau, 
 with $Q \to K_*$)
\eqn\teem{ T\sim {a^{1/4} 
K_*^{-1/2}} \ .
}
We should express the temperature in terms of the glueball mass scale \KS 
\eqn\laam{ \Lambda = L_{\rm P} P^{-1}  \ , 
}
which we also expect to be the scale of the critical temperature: 
$T_c \sim \Lambda$.
Using \laam\ and \hihi\ we find
\eqn\ttt{ T
\sim \ \Lambda\  \sqrt \l \  e^{ 1/( 2\l)} \sim T_c \sqrt \l \  e^{ 1/( 2\l)}\ 
\ .
}
Thus, in our perturbative regime we find $T\gg T_c$; as expected, 
this regime is 
applicable far above the phase transition into the chiral symmetry restored
phase.

\subsec{\bf Solutions for other fields }

Let us   now solve for perturbations of other fields.
Using \aau\  and \ffe\  the  equation for  the dilaton
 \pu\  becomes 
\eqn\poio{
\p''  = { 4 a^2 \ov K_*} { 1 
- e^{-8a u} \cosh^2 4ak \ov \sinh 4 a (u +k)\  \sinh 4 a u }
\ .}
Expanding in powers of $u$ 
 this gives
$\p''  = - {  4 a^2 \ov K_*^2 u }  + c_1 + c_2 u + .. .$, 
so  that the small $u$ asymptotics of 
$\p$ is $u \log u$.
To simplify the expressions, let us consider 
again the case of $k=0$. Then 
\eqn\piol{
\p''  = { 16 a^2 \ov K_* ( e^{8a u} -1 )}
\ , \  \ \ \  {\rm i.e.} \ \ \ \ \ 
 \p'=  {  2 a \ov 2 K_* } \log (1 - e^{ - 8a u} ) 
\ , } 
where we  have fixed the integration constant so that 
$\p'\to 0$ at $u\to \infty$,  
\eqn\poll{
\p= \p_*  
+   { 1 \ov 4 K_*} \Li_2(e^{-8au}) \ ,\ \ \ \ \ \  \ \ \ \ 
\p_* = -{\pi^2\over 24 K_*} \ . }
$\Li_n(z)$ denotes the polylogarithm function.
The limits of this expression are 
\eqn\smalu{
u\to 0:\ \ \ \ \  \p=  
  {  2 a \ov K_* }  [ u( \log 8au  - 1)  
    + O( u^2 \log u)  ] \ ,   }
\eqn\loar{
u\to \infty: \ \ \  \  \ \ 
\p =\p_*  + { 1 \ov 4 K_* } e^{-8au} + O(e^{-16au})
\ .  }
This implies that the string coupling $e^\Phi$  
decreases as we approach the horizon.

Next, we  are to find $\om,\xi,\eta$
 satisfying \yyu,\typs,\zy, i.e. (for $k=0$) 
\eqn\typk{ 
\om''  -  { 12 a^2 \ov \sinh^2 4 au }  \om   =
{ 8  a^2 \ov  5 K_* (e^{8a u} -1) }
\ ,  }
\eqn\xxx{
 \xi''  - { 32 a^2 \ov \sinh^2 4a u} 
   \xi     =  -  { 8 a^2 \ov  5K_* ( e^{8a u} -1 )}
\ , }
\eqn\zzz{
\eta''  -  { 32 a^2 \ov \sinh^2 4a u} 
   \eta    
=  {  a^2 \ov K_* \sinh^2 4 au }
[1 +   e^{- 8 a u}  - 4 \log (1 - e^{-8au})    ] 
\ .   }
To  analyze these equations for $a\not=0$ 
it is convenient 
to introduce a new radial variable 
\eqn\vev{ v = 1-e^{-8au} \ .   } 
Then \typk, \xxx, and \zzz\ can be expressed in the
forms
  \eqn\typkH{
   v(1-v) \om'' - v \om' - {3/4 \over v} \om 
 = {1 \over 40 K_*} \ ,
  }
  \eqn\xxxH{
   v(1-v) \xi'' - v \xi' - {2 \over v} \xi = -{1 \over 40 K_*} \ ,
  }
  \eqn\zzzH{
   v(1-v) \eta'' - v \eta' - {2 \over v} \eta = 
    {1 \over 16 K_* v} (2-v-4\log v) \ ,
  }
 where primes now denote $d/dv$.
  Now, the homogenous equation 
  $$v(1-v)
f'' - v f' - {A \over v} f = 0
$$ 
is solved for generic $A$ by $f(v) =
v^\nu  \ {}_2F_1(\nu,\nu;2\nu;v)$,
 where ${}_2F_1$ is the hypergeometric function and $\nu(\nu-1)=A$.
As it happens, $A=2$ is a degenerate case where the solutions to the
homogenous equation are elementary functions of $v$ (namely, ${1 \over
v} - {1 \over 2}$ and $-2 + {v-2\over v} \log (1-v)$).  Using these
solutions one can extract solutions to the inhomogenous equations as
well:
  \eqn\xxxSoln{
   \xi = {2v + [-2v+(v-2) \log(1-v)] \log v + 
    (v-2) \Li_2(v) \over 40 K_* v} \ ,
  }
  \eqn\zzzSoln{
   \eta = {v-2 \over 16 K_* v} \left[ \log v
   \log (1-v) + \Li_2(v) \right] \ .
  }
For $v\to 0$, which corresponds to $u\to 0$, we have the asymptotics
  \eqn\xiAsymptZero{
   \xi \sim {v \over 80 K_* } + O(v^2 \log v) \ ,
  }
  \eqn\etaAsymptZero{
   \eta \sim {\log v - 1 \over 8 K_*} + {v \over 32 K_*} + O(v^2 \log v) \ . 
  }
 This changes the AdS asymptotics into the KT asymptotics in agreement with
 \lla.
 
For $v \to 1$, which corresponds to $u\to \infty$, we have 
the near-horizon asymptotics  consistent with 
our expectations \boun,\bam:
  \eqn\xiAsymptOne{
   \xi \sim {12-\pi^2 \over 240 K_*} + {9-\pi^2\over 120 K_*}(1-v)
   + O[(1-v)^2] \ ,
  }
  \eqn\etaAsymptOne{
   \eta \sim -{\pi^2 \over 96 K_*} + {3- \pi^2\over 48 K_*}(1-v)
 + O[(1-v)^2]\ . }
To summarize, the horizon values of the perturbations we have solved
for are
\eqn\horval{
\xi_* = {12-\pi^2 \over 240 K_*}\ , \quad\ \ \ 
\eta_* = -{\pi^2 \over 96 K_*}\ , \quad\ \ \ 
\phi_* = -{\pi^2 \over 24 K_*}\ .
}

The ``squash factor'' $\om$  in \mmm\ is special in that the volume of
compact space does not depend on it. Hence it cancels from
the observables like the horizon area and the Hawking temperature.
An explicit expression for $\omega$ is also available in principle,
but it involves some messy integrals, so we will just determine its 
asymptotics schematically. 
At large $u$ \typk\ becomes
\eqn\typy{ 
\om''  -  { 48 a^2 e^{-8 au }}  \om   =
{ 8 a^2 \ov 5 K_* } e^{-8a u}
\ ,  }
and  a solution  of this  equation is 
given in terms of Bessel functions of $e^{-8au}$
so that the asymptotics is 
\eqn\sop{
\om=  \om_* + \om_ 1 e^{-8a u} + O( e^{-16a u}) + ... \ , \ \ \ \ \ \ \ 
\om_1 = {3\over 4} \om_* + {1\over 40 K_*}\ . }
For small $u$  (again, in the $k=0$ case)
we get
\eqn\typi{ 
\om''  -  { 3 \ov 4 u^2 }  \om   =
{  a \ov 5 K_* u }
\ ,  }
with the  general  solution  
satisfying \lla\  being 
\eqn\goo{
u\to 0: \ \ \ \ \ \ \ \ \  \om
= -{  4 a \ov 15 K_* } u   + b_1 u^{ 3/2   }  
\ . }

Entropy and temperature can be determined straightforwardly now that
the metric is known.  
The reader will have no trouble
verifying that the metric \mett\ 
can be cast into the form
$$     ds_{10E}^2 = 
{\sqrt{ 8 a/K_*}\ov  \sqrt{v}}  e^{2P^2 \eta}
\left[ (1-v) 
    dX_0^2 +  dX_i^2 \right] + 
    {\sqrt{K_*} \over 32} e^{-2P^2 (\eta -5\xi)}
    {dv^2 \over v^2 (1-v)}  $$
 \eqn\FoundMet{
 + \ 
    {\sqrt{K_*} \over 2} e^{-2P^2 (\eta-\xi)} \left[
      e^{-8P^2 \omega} e_\psi^2 + 
     e^{2P^2 \omega} (e_{\theta_1}^2 + e_{\phi_1}^2 + 
      e_{\theta_2}^2 + e_{\phi_2}^2) \right] \ .
  }
 Using the explicit formulas for $\xi$ \xiAsymptZero\
and $\eta$ \etaAsymptZero, we obtain an
explicit expression for the entropy per unit volume divided by the
temperature cubed:
  \eqn\EntTempRatio{
   {S \over VT^3} = \alpha {K_*^2\over L_{\rm P}^8}
    e^{4 P^2(5 \xi_* - 2 \eta_*)}
    =\alpha  {K_*^2 \over L_{\rm P}^8}
    \left[ 1 + \lambda + O(\lambda^2) \right]
 \ ,  }
 where $\alpha $ is a factor of order unity which can be fixed by
specializing to the $P=0$ case and using the results of \gkp.
Unravelling the various definitions that we have made, one can show
that $K_*/L_{\rm P}^4$ is the flux of the five-form through $T^{1,1}$,
measured in Dirac units.  This is the effective number of colors
available at a given scale.  (As an aside, it is gratifying to observe
that the many factors of $2$, $3$, and $\pi$ cancel out to give a
simple $1$ as the coefficient of $\lambda= P^2/K_*$ in the function in
square brackets.)

The formula \EntTempRatio\ does not identify how $K_*$ depends on $T$.
To determine this, one must be careful in defining exactly what
temperature means in this non-asymptotically flat geometry.
Equivalently, we need to specify a way of normalizing time which is
invariant under changes of $a$.  A sensible procedure is to pick some
very large $K_0$ (much larger than $K_*$ for the range of $a$ we are
considering) and then define a privileged time coordinate $\tau$ by
the condition  that $g_{\tau\tau} = 1$ at the radius where when $K(u) = K_0$.
One may easily verify from \FoundMet\ that the coordinate we have
called $X_0$ is {\it not} such a coordinate.  But it is equivalent to
compute the Hawking temperature from \FoundMet\ in the usual way (by
eliminating the conical deficit at $v=1$) and then multiply by the
appropriate power of $g_{00}$ to convert to what one would have found
using $\tau$.  The end result is precisely \ttt, which can be inverted
to read
  \eqn\tttInverse{
   {K_* \over 2P^2} = \log {T \over \Lambda} + 
     {1 \over 2} \log\log {T \over \Lambda} + \ldots \ .
  }
 Recall that $\Lambda$ is the scale of glueball masses.  Neglecting
the $\log\log$ correction, we recover a result expected by Buchel
\Buch:
  \eqn\BuchRatio{
   {S \over VT^3} \sim {P^4 \over L_{\rm P}^8} 
    \left( \log {T \over \Lambda} \right)^2 \ .
  }
 Physically speaking, we discover that the effective number of colors
at a given temperature $T$ rises as $\log (T/\Lambda)$.  
This is consistent with
the finding \krasnitz\ that two-point
functions,  computed from supergravity,  scale as a negative power of the
separation $x$ times $(\log x \Lambda)^2$.

On the
supergravity side, it seems reasonable to suppose that the critical
solution has $K_* \to 0$ (that is, zero $F_5$ at the horizon), since
this is where, as suggested by \EntTempRatio,
  the entropy must go to
zero -- and we believe it must do so in order for correlators to have
power law tails.  All this is consistent with the view that chiral
symmetry breaking occurs simultaneously with confinement in the gauge
theory.

\newsec{Concluding Remarks}

We have carried out a perturbation expansion in $P$ to
the leading non-trivial order for all the fields involved in the
non-extremal KT solution.  
The real expansion parameter turns out to be $\lambda =
P^2/K_*$.  Since $K_*$ is the five-form flux at the horizon, and this
quantity gets bigger and bigger as we push the horizon further into
the UV, it is clear that the small $\lambda$ expansion is precisely a
high temperature expansion (see \ttt).  At the first non-trivial order in
perturbation theory, we find a smooth interpolation between the KT
solution and the near-extremal D3-brane -- or,  to put it differently, a
regular non-extremal generalization of the KT geometry.  It is
necessary for the $U(1)$ fiber of $T^{1,1}$ to become squashed, and for
the dilaton to run.  There is a mild pathology in the perturbation
theory, in that both $F$ and $\eta$  in \seeq,\ded\ acquired $\log u$  divergences
in the UV; but these are precisely the divergences needed to match
onto the KT solution.  The perturbation 
 fields approach constant values at the IR 
horizon, which guarantees that this is a regular horizon, not a
singularity.  Further terms in the perturbation series should be {\it
uniformly} small for all fields.
We conclude that the supergravity dual of the
$SU(N+M)\times SU(N)$ gauge theory at very high temperature
is described by a background containing a well-developed
Schwarzschild horizon in an asymptotically KT geometry, as suggested in
\Buch.
This background is $U(1)$ symmetric; therefore, the chiral symmetry is
indeed restored at high enough temperature. 

As we increase the value of $\l=P^2 K_*^{-1}$,
which corresponds to decreasing the temperature,
the perturbation theory
becomes less reliable.  The obvious alternative is numerical
integration.  The main difficulty here is that some of the
boundary conditions on the second order differential equations are set
at the horizon, while the others are set in the UV by requiring KT
asymptotics.  This makes ``shooting'' algorithms tedious since one is
operating in a multi-dimensional space of boundary conditions.  It is
amusing that the perturbative approach circumvents this difficulty,
and it is possible that some hybrid of 
 a perturbative treatment
and numerics might be developed.
There is a good motivation for studying the temperature dependence of
the background: it might be possible to approach the chiral symmetry
breaking
point $T=T_c$ and characterize the dual of the critical theory.


\bigskip
\noindent
{\bf Acknowledgements}
\bigskip
We are grateful to L. Pando Zayas for collaboration at early stages of
this project.  The work of S.~S.~G.\ was supported in part by the DOE
grant DE-FG03-92ER40701.
The work of C.~P.~H. was supported in part by the DoD.
The work of I.~R.~K. was supported in part by the NSF
grant PHY-9802484.
The work of A.~T. is supported in part by
the DOE grant  DE-FG02-91ER-40690, PPARC SPG grant,
and INTAS project 991590.

\appendix{A}{\bf Asymptotic Expansions of the Solution}

In this Appendix we consider the asymptotic solutions of our system in
the near-horizon (large $u$) and long-distance (small $u$) limits.  In
the absence of exact solutions, this a necessary preliminary to any
approximation scheme, numerical or otherwise, for obtaining a complete
solution.

To gain some intuition about the role of $w$ we shall start in
section~A.1 with an analysis of the solution in the absence of any
matter  fields.  Thus only $y''$ and $w''$ will be  non-zero.  Then in
section~A.2 we shall treat the full system of equations.  
To simplify notation, we work in units where $L_{\rm P}=1$.

\subsec{\bf  Solutions  for 
  $y$  and $w$    in the 
absence of matter}
Assuming $Q=P=f=\P=0$ and  choosing $x$  and $z$ to satisfy \boun, 
 the 2-nd order system 
\yyy--\coln\ reduces to 
\eqn\copi{  x'=a\ , \ \ \  \ \ \ \ \ z'= -a \ ,    }   
\eqn\subs{
y'' =    { 4 \ov 5} e^{8y} (6 e^{-2w} - e^{-12 w}) \ , \ \ \ \ 
w'' = { 6\ov 5}     e^{8y} ( e^{-2w} - e^{-12 w})  \ , } 
\eqn\lop{  y'^2  -  w'^2  
 -   {1 \ov 5}  e^{8y} ( 6 e^{-2w} - e^{-12 w} )   = a^2 \ . 
}
If the constraint  \lop\ 
is satisfied at one point $u=u_0$, it is satisfied for all $u$
because of the equations of motion \subs.  We can satisfy
the constraint at $u\to \infty$ by choosing the asymptotics to 
be as in \boun\
\eqn\ops{ 
\ y'\to -a\ , \ \ \  \ \ \ \ \  \ \ 
 w'\to 0  \ .    }   
Let us  first set $a=0$ and 
consider the extremal BPS  solution 
described by the 1-st order system 
$y' + { 1 \ov 5} e^{4y}(3 e^{4 w} + 2   e^{-6w} )  
     =0, $ \ \  $
w' - { 3 \ov 5} e^{4y} (e^{4w} -  e^{-6 w})    
   =0 $ (see equation (3.11) of \refs{\bhkpt} and also \refs{\KT,\PTtwo}).
{}From the equation for $dy\ov dw$ one finds that 
\eqn\soll{
  y=  -  \int dw  {3 e^{ 10 w} + 2   \ov 3 (e^{ 10 w}  - 1 )  }
= y_0  - { 1 \ov 6} \log  ( e^{6 w} -e^{ - 4w})  \ ,  }
i.e. 
\eqn\kik{
e^{-6(y-y_0)} = e^{6w} - e^{-4w} \ . } 
That gives
$
w' ={ 3\ov 5} e^{4 y_0} (1-  e^{-10 w})^{1/3} , 
$
implying that $w$ can be expressed implicitly, $u=u(w)$, in terms of a sum 
of logarithms and arctangents.  We set the integration constant by requiring
that $w=0$ when $u=0$.
Introducing  $\r= e^{y+w}$,  the  resulting  
  Ricci flat 6-d metric  is then  the  generalized conifold
of \PT\
\eqn\genn{
ds_6^2={ \kappa^{-1}(\r)}d\r^2+  \r^2 \big[ 
\kappa (\r) 
e_{\psi}^2 +  e_{\theta_1}^2+e_{\phi_1}^2+
e_{\theta_2}^2+e_{\phi_2}^2\big] ,
}
\eqn\kaa{
\kappa(\r)=  e^{-10w} = 1-{\r_*^6 \over \r^6} \ , \ \ \ \  \ \ \
\r = e^{ y +  w } , \ \ \ \     \r_* = e^{y_0}
  \leq \r < \infty  \ . }
  $w$ changes  from 0 at $\r=\infty$ ($u=0$)  to $\infty$ at 
$\r=\r_*$ ($u=\infty$), while  
$y=\infty$ at $\r=\infty$ and $y= -\infty$ at $\r=\r_*$.
The asymptotics near 
$\r=\r_*$, i.e. $u\to \infty$ are  
\eqn\asyy{
y=- c u +  y_*   +  y_1 e^{-10 c u}+  ... \ , \ \ \ \
w= c u +  w_* + w_1  e^{-10 c u}+ ..  ... \ , 
 }
where
\eqn\asyyconst{
c  = { 3 \ov 5} e^{4 y_0} \ , \ \ \ \
 \ \ \ \  y_*= y_0 - w_* \  , \ \ \ \ \ \ 
w_* = {\sqrt{3} \pi \ov 60} - {3 \ov 20} \log 3 \ ,}
and  $
y_1 = {2 \ov 15} e^{-10w_*} , \ \ 
w_1 = {1\ov 30} e^{-10w_*}  . $
To find the ``non-extremal" analog of this space, we are to 
switch on $a\not=0$. Then  the system \subs, \lop\
does not have a simple analytic solution,
but it is easy to study the asymptotics at small and large $u$.
The long-distance
asymptotic solution is  
\eqn\largd{
u\to 0: \ \ \ \ \ \  y =  - { 1 \ov 4} \log 4 u  
  - { 2 \ov 3} a^2 u^2  - { 3\ov 4 } s^2 u^3  +
   { 16\ov 45 } a^4 u^4  + {16\ov 11}   s^3  u^{9/2}  +  ... 
\ , }
\eqn\wla{
w =     s  u^{3/2}   -  s^2  u^{3}  - {1\ov 2 } a^2 s u^{7/2} +
 {11\ov 6}  s^3  u^{9/2}   + ... 
 \ .   }
 This is  consistent with the extremal generalized 
conifold solution \kaa\ 
in the $a\to 0$ limit  if  
$ s= { 4 \ov 5} \r_*^6 $.
Taking instead  $s$ to  be proportional to $a$ 
 we    recover   the conifold solution  with $w=0$ in the $a\to 0$ limit.


The short-distance asymptotics is described by 
\eqn\lagg{
u\to \infty: \ \ \ \ 
y=- b u +  y_*  +  y_1^{(1)} e^{- (8b + 2\n)  u} + y_1^{(2)} 
e^{-  (8b + 12 \n)  u} +  O (e^{- 2(8b + 2\n)  u})  \ ,}
\eqn\jio{ 
w=     \n u +  w_* +   w_1^{(1)} e^{- ( 8b + 2\n)  u} +  w_1^{(2)} 
e^{- (8b + 12 \n) u}  +  O (e^{- 2(8b + 2\n)  u})   \ , 
}
where  
\eqn\coop{
b^2 - \n^2 =a^2  \ , \ \ \ \   \ \ \ 
y_1^{(1)}  = {24 e^{8y_*-2w_*} \ov 5 (8b + 2\n)^2 } \ , \ \ \ 
w_1^{(1)}  = { 6  e^{8y_*-2w_*}\ov 5 (8b + 2\n)^2  }  \ ,    
}
\eqn\cooptwo{
y_1^{(2)} = -{4 e^{8y_*-12w_*} \ov 5 (8b + 12\n)^2} \ , \ \ \ \
w_1^{(2)} = -{6 e^{8y_*-12w_*} \ov 5 (8b + 12\n)^2}\ .
}
These asymptotics reduce to the extremal solution
\asyy\  in the limit $a\to 0$ provided we set $b = c$. 

A numerical analysis confirms the existence of a solution of \copi--\lop\  interpolating between the two asymptotic regions 
\largd, \wla\ and \lagg, \jio.

 To have a solution reducing  to the standard ($w=0$) 
conifold one  in the $a\to 0$ limit, 
we should set\foot{One  is 
 to resum the series in \lagg\ in order 
 to reproduce the 
exact conifold solution $y= -{ 1 \ov 4} \log 4u $.} 
\eqn\sett{
\n=  n a \ , \ \ \ \ \   b=  (1 + n^2)^{1/2}   a\ , \ \ \ \ \ \  e^{4y_0} 
 = {5 \ov 3} b  
\  .   }
To satisfy \ops\ we are to assume the simplest possibility --
$n=0, \n=0$, i.e. $ b=a$. 
For $\n=0$  the function $w$ goes  to a constant   instead 
of infinity  and  the asymtotic  $u \to \infty$ solution is 
found to be 
\eqn\lagge{
u\to \infty: \ \ \ \ \ \ 
y= - a u + y_*  +  y_1 e^{-8au} + ... \ , \ \ \ \ \ \
y_1 = { 1 \ov 80 a^2}  e^{8y_*}
(6 e^{-2w_*} - e^{-12w_*}) \ , }
\eqn\kio{
w = w_* + w_1  e^{ - 8 a u} + ...  \ , \ \ \ 
\ \ \ \ \ \ \  w_1 = { 3 \ov 160 a^2} e^{8y_*} 
(e^{-2w_*} - e^{-12w_*}) \ .  
}
The behaviour of $w$  follows from  the equation
$w'' =  12  e^{8y_*  - 8au}  w + O(w^2)$.
Once we include charges, this equation
(i.e.  \yuw)  will have an extra inhomogeneous $O(a^2 P^2)$ term (cf. \typs,\piol).
  The variation of 
$w$  should  be driven only by the non-extremality
 {\it and } non-zero $P$, 
so that  $w$ should  vanish in any of the limits
 $a\to 0$ or $P\to 0$.

For finite $a$ the point $u=\infty$ is the horizon.
For general $\n$  (i.e. $b$ not 
necessarily equal to $a$)
 the near-horizon metric \mett\
 becomes (using   \copi,\lagg,\jio):
$$ds^2_{10E}\ \ \  \to  \  \  e^{2z_* } ( e^{-8au} dX_0^2 +  dX_i dX_i) 
+  e^{ 10 y_* -2z_* - 10 b u + 2 a u} du^2  
$$\eqn\mety{
+ \    e^{2y_* -2z_*  - 2 b u  +  2 au} \big[
e^{ -8w_* - 8 \n  u   }  e_{\psi}^2 
+  e^{ 2w_* + 2\n u   }
\big(e_{\theta_1}^2+e_{\phi_1}^2 +
e_{\theta_2}^2+e_{\phi_2}^2\big)\big] \ . }
While the Ricci tensor of this metric
 vanishes at large $u$, 
the   full curvature   invariants 
are singular at $u=\infty$ 
 for generic values of $\n$.\foot{For example, 
choosing $\n=a$ we get
the  leading singularity as 
$ (R_{....})^2 \sim  e^{ \kappa  u}$,
where $\kappa= 4(5\sqrt 2-1) > 0$. }

The {\it only}  case when the horizon is {\it regular}
 is precisely the one 
we are interested in: 
$\n=0, \ b=a$. In this case  \mety\ 
is indeed  the asymptotic  form of this  ``black hole on conifold", i.e.
 the   zero-charge  ($q=0$, $h=1$) case of the 
regular non-extremal D3-brane  metric \hji,\qqq.
In this case \mety\ factorizes into 
$R^2_{u,X_0}\times R^3_{X_i} \times T^{1,1}$ with the  flat 
$ (u,X_0)$   part being 
\eqn\paa{
ds^2_2 = {  e^{10 y_*-2z_* } \ov 16 a^2} (dU^2 +  
 U^2  d\td X_0^2) \ , \ \ \   
\  U=  e^{- 4 au}\  , \ \ \ \ 
 \td X_0 =  4a  e^{2z_* -5  y_*} X_0 \ . }
 As usual,  $X_0$ is 
to be  made periodic to avoid the conical singularity, thus determining
the Hawking 
temperature \haw\ of this Schwarzschild  7-d  black hole.

It is this $b=a, \ \n=0$ solution  with 
non-vanishing $w$ 
interpolating between  \largd, \wla\ and \lagge\ 
that we  would like now 
to generalize to the  fractional D3-brane case.

\subsec{\bf Asymptotic solutions  for  non-zero charges  }

We shall proceed 
by determining the asymptotic  $u\to \infty $ and $u\to 0$ 
solutions 
of the system \yyy--\coln,  now  for non-vanishing charges.
It is the non-extremality parameter $a$ 
that should drive the variation of $w$ and $\P$, 
so we should choose  the integration 
constants  appropriately.

Let us start with the near-horizon region.
Generalizing \boun,\lagge,\kio\ we 
shall set
\eqn\bun{
 \ u\to \infty  : \  \ \
x = a u \ , \ \ \ \   y\to  - a u  + y_* + y_1 e^{-8au}  + ...\ , 
\ \ \  \  z\to  - au +  z_* +  z_1  e^{-8au}  + ... \ ,}
\eqn\bunt{ w\to w_*  + w_1 e^{-8au} + ...\ , \ \ \ \ \
 \P\to \P_* + \P_1 e^{-8au} + ...\ , 
  \ \ \ \  K\to  K_* + 2P f_1 e^{-8au} + ...
\ ,    }
where the expansion goes in powers of $e^{-8au}$.
These asymptotics are  justified also  by 
the results of   perturbation theory in $P$
 starting with the 
near-horizon region of the standard  
non-extremal D3-brane as described in section 5. 
(Note that $K_*$ is the same as in \seeq\ up to terms of $O(P^2)$.)

  It is straightforward to check that this ansatz is consistent 
with our full system of equations \yyy--\coln, determining the coefficients as follows (cf. \lagge,\kio):
\eqn\coe{
640 a^2 y_1= 8   e^{ 8y_*}(6 e^{ -2 w_*} -
e^{  -12 w_*})
- P^2  e^{ \P_* + 4y_*  + 4z_* +  4 w_*} \ , }
\eqn\coew{
640 a^2  w_1= 12   e^{ 8y_*}( e^{ -2 w_*} -
e^{  -12 w_*})
+ P^2  e^{ \P_* + 4y_*  + 4z_* +  4 w_*} \ , }
\eqn\coez{
256  a^2 z_1 =   K_*^2  e^{ 8z_* } 
+ P^2  e^{ \P_* + 4y_*  + 4z_* +  4 w_*} \ ,
}  
\eqn\coef{
64 a^2 f_1= P K_*  e^{ \P_* + 4y_*  + 4z_* +  4 w_*} \ , \ \ \ \ \ \ \ 
64 a^2 \P_1  = P^2 e^{ \P_* + 4y_*  + 4z_* +  4 w_*} \ . }
In order to satisfy the  requirement 
of correspondence with 
the  D3-brane  solution  in the  $P\to 0$ limit  
 we  may   choose
 \eqn\seti{
e^{4y_*} = 2 a e^{4P^2 \xi_*}\ , \ \ \ \   \ \ 
  e^{4z_*} = { 8a\ov  K_*} e^{4P^2 \eta_*} \  , \ \  \ \ \ \ \ 
e^{w_* } = e^{P^2\omega_*} \ , \ \ \ \ 
e^{\P_*} = e^{P^2 \phi_*} \ .     }
Then, to $O(P^2)$, \coe--\coef\ become 
\eqn\coee{
y_1=   { 1 \ov 4} 
+ { P^2  \ov 120 K_*}(9-\pi^2)    \ ,  \ \ \ \ \ \ \ 
 w_1=   { P^2  \ov 40 K_*}(1+30\omega_* K_*)  \ , }
\eqn\cez{
  z_1 =     { 1\ov 4 }    +
 { P^2  \ov 48 K_*}(3-\pi^2)  \ ,  \ \ \ \ \ 
 f_1= { P \ov  4}   \ , \ \ \ \ \ \ \ 
\P_1  = { P^2  \ov  4 K_*} \ . } 
The values of these coefficients 
are in agreement  with the  large $u$ 
 limits of the leading-order in $P$  corrections to our 
fields found in section 5
(see \samolet,\loar,\xiAsymptOne,\etaAsymptOne\ and \sop).

Next, we need to fix the asymptotic  behaviour  at $u\to 0$. 
We should reproduce  
the  black D3-brane asymptotics \sss,\sis\ 
for $P=0$,  and  the  KT asymptotics \kot,\ktt\ for $a=0$.
It is therefore reasonable to assume that the first terms in the asymptotic 
expansion are (in the $k=0$ case in \sis) 
\eqn\ggg{
y= - { 1 \ov 4} \log 4 u + ...
\ ,  \ \ \ \
e^{-4z} =   u( \z_{1,0} + \z_{1,1}  \log u )
+ ... \ , \ \ \ \
K=  - { P^2\ov 2 } \log u + ... \ .  }
Moreover, we expect $w$ and $\Phi$ to vanish at $u=0$.
It turns out to be easier to find expansions for $\exp(-4w)$, $\exp(-2w)$, 
and $\exp(-\Phi)$ than for their corresponding logarithms.  
We have found a small $u$ expansion of the following form  
(assuming  again that the asymptotically flat region is omitted, i.e. $k=0$):
\eqnn\Yu
\eqnn\Zu
\eqnn\Ku
\eqnn\Wu
\eqnn\Eu
$$
\eqalignno{
e^{-4y}  &= 4u + \y_1 u^2 + 
u^3 (\y_{2,0} + \y_{2,1} \log u + \y_{2,2} (\log u)^2) + \y_3 u^{7/2}
	+ O(u^4) \ , &\Yu\cr
e^{-4z}  &= u(\z_{1,0} + \z_{1,1} \log u ) + u^2 (\z_{2,0} + \z_{2,1} \log u )
       + \z_3 u^{5/2}   \cr
	& \ \ \ +  u^3 (\z_{4,0} + \z_{4,1} \log u + \z_{4,2} (\log u)^2 + \z_{4,3} (\log u)^3)
          + \z_5 u^{7/2} + O(u^4)  
\ , &\Zu\cr
K &= 
- {1\ov 2} P^2 \log u + 2P \bigg( u(\f_{1,0} + \f_{1,1} \log u ) + \f_2 u^{3/2}  \cr 
	& \ \ \
	+ u^2 ( \f_{3,0}  	
 + \f_{3,1} \log u +  \f_{3,2} (\log u)^2)+
	u^{5/2} (\f_{4,0} + \f_{4,1} \log u )
	+O(u^3) \bigg) \ , &\Ku\cr
e^{-2w} &= 1 + \w_1 u + \w_2 u^{3/2} + 
u^2( \w_{3,0} + \w_{3,1} \log u ) + \w_4 u^{5/2} + O(u^3)
\ , &\Wu\cr
e^{-\Phi} &= 1 + u(p_{1,0} + p_{1,1} \log u ) + u^2(p_{2,0} + p_{2,1} \log u + 
p_{2,2} (\log u)^2 ) \cr
	& \ \ \
       + p_3 u^{5/2} + O(u^3) \ , &\Eu\cr}
$$
It is easy to write the relations between the first few coefficients:
\eqn\urelations{
\z_{1,1} = -{1\ov 2} P^2 \ , \ \ \ \
\z_{1,0} =  {1\ov 2}P^2  \ , \ \ \ \
p_{1,1} = {5\ov 4} \y_1 \ , \ \ \ \ \w_1 = -{1\ov 3} \y_1 \ .
}
Certain other coefficients are also relatively simple:
$$
\y_{2,2} = {5\ov 6} \y_1^2 \ , \ \ \ \
\y_3 = -{48 \ov 7P} \y_1 \f_2 \ , \ \ \ \
\z_{2,1} = {5\ov 8} P^2 \y_1 \ , \ \ \ \
\z_3 =  {4\ov 5} P \f_2 \ , 
$$
$$
\z_{4,3} = -{25\ov 192} P^2 \y_1^2 \ , \ \ \ \
\f_{1,1} = {5\ov 8} P \y_1 \ , \ \ \ \
\f_{3,2} = -{35\ov 96} P \y_1^2 \ , \ \ \ \
\w_2 = {3\ov P} \f_2 \ , \ \ \ \
$$
\eqn\urelationsmore{
\w_{3,1} = -{1\ov 2} \y_1^2 \ , \ \ \ \
\w_4 = -{29\ov 8P} \y_1 \f_2 \ , \ \ \ \
p_{2,2} = {25\ov 32} \y_1^2 \ , \ \ \ \
p_3 = -{2\ov P} \y_1 \f_2 \ .
}
The remaining coefficients are more 
complicated, and we will not write them down.
All of the coefficients can
be expressed in terms of 
$\y_1$, $\f_{1,0}$, $\f_2$, $\y_3$, and $\f_{3,0}$.  
In other words, these five coefficients correspond to
five undetermined integration constants.

The relations 
\urelations\  between 
$\y_1$, $p_{1,1}$ and $\w_1$ are
consistent with the small $P$ expansion results of section 5 
(see \samolet, \smalu, \xiAsymptZero, \goo):
\eqn\agreq{
y = -{1\ov 4} \log(4u) - {1\ov 16} \y_1 u + ... \ , 
\ \ \ \ \ \ 
K(u) =  - {P^2\ov 2} \log u + 2P \f_{1,0} u + ... \ ,} 
\eqn\zoiks{
w = {1\ov 6} \y_1 u + ... \ , \ \ \ \ \ \   \ \ \
\Phi = -{5\ov 4} \y_1 u \log u + ... \ . 
}
In order to find perfect agreement, we must set
\eqn\agreement{
K_* =  {P^2\ov 2} \log(8a) + O(P^4) \ , \ \ \ \
\f_{1,0} = a P \ , \ \ \ \ \y_1 = -{8 a P^2 \ov 5 K_*} \ .
}
Given these values and the relation
$\z_{2,0} = P \f_{1,0} - {5\ov 16} P^2 \y_1$,
one may also check that the expansion
\Zu\ is consistent with \zzzSoln. 
To sum up, 
this comparison with the small $P$ expansion has
allowed us to fix two of the integration constants to
$O(P^2)$, and presumably still more information could be
extracted.

There are some differences between the small $u$ expansion 
and the expressions found from perturbing in $P$.  For example,
from \samolet\ one would expect $\f_{1,1}=0$ while from
\urelationsmore\ it is clearly not.
However, $\f_{1,1}$ is $O(P^3)$
and hence would not be expected to show up at the order $P^2$
calculated in section 5.

Thus far, we have not considered the constraint equation
\coln.  We have calculated the expansion to one more order in
$u$ than was needed to extract useful information from the constraint.  The
constraint gives one  
relation between $a$, $\f_{1,0}$, $\y_1$, $\y_3$, and
$\f_{3,0}$   which is  complicated
and is not  reproduced here.  The important
point is that we have enough freedom to choose
$\f_{1,0}$ and $\y_1$ as in \agreement\ and still satisfy the constraint.

The boundary conditions found above are the starting point 
for a numerical analysis which is to demonstrate 
that the $u\to 0$ and $u\to \infty$ asymptotics can be smoothly connected.


\vfill\eject
\listrefs
\end